\documentclass[aps,pra,reprint,amsmath,amssymb,floatfix,citeautoscript,superscriptaddress]{revtex4-1}
\usepackage{amsmath}
\usepackage{graphicx}
\usepackage[usenames,dvipsnames]{xcolor}
\definecolor{myblue}{rgb}{0,0,1}
\usepackage[colorlinks=true,linkcolor=myblue,citecolor=myblue]{hyperref}

\usepackage[T1]{fontenc}
\usepackage{txfonts}

\def\b{\textbf}
\def\t{\textrm}

\begin{document}

\title{Binding energies and spatial structures of small carrier complexes\\
in monolayer transition metal dichalcogenides via diffusion Monte Carlo}

\author{Matthew Z. Mayers}
\affiliation{Department of Chemistry, Columbia University, New York, NY 10027, USA}
\author{Timothy C. Berkelbach}
\affiliation{Princeton Center for Theoretical Science, Princeton University, Princeton, NJ 08544, USA}
\author{Mark S. Hybertsen}
\affiliation{Center for Functional Nanomaterials, Brookhaven National Laboratory, Upton, NY 11973-5000, USA}
\author{David R. Reichman}
\affiliation{Department of Chemistry, Columbia University, New York, NY 10027, USA}

\begin{abstract}
Ground state diffusion Monte Carlo is used to investigate the binding energies
and carrier probability distributions of excitons, trions, and biexcitons in a
variety of two-dimensional transition metal dichalcogenide materials. We
compare these results to approximate variational calculations, as well as to
analogous Monte Carlo calculations performed with simplified carrier
interaction potentials. Our results highlight the successes and failures of
approximate approaches as well as the physical features that determine the
stability of small carrier complexes in monolayer transition metal
dichalcogenide materials. Lastly, we discuss points of agreement and
disagreement with recent experiments.
\end{abstract}

\maketitle

Atomically thin layers of crystalline transition metal dichalcogenides (TMDCs)
have been the subject of intense investigation in recent
years~\cite{Splendiani:19,Mak:20}.  As with graphene, TMDCs exhibit remarkable
properties that originate from their quasi two-dimensional
nature~\cite{Butler:21,Geim:22}.  However unlike graphene, TMDCs are direct gap
semiconductors, opening up a wealth of potential practical applications ranging
from field-effect transistors to photovoltaics~\cite{Chhowalla:23,Jariwala:24}.
Furthermore, due to the lack of inversion symmetry in these single layer
crystals, the so-called $K$-points on opposite corners of the two-dimensional
hexagonal Brillouin zone are inequivalent~\cite{Li:25,Lebegue:26,Gunawan:27}.
As a result, a distinct valley degree of freedom associated to states near
these points emerges which may be manipulated and controlled, leading to the
possibility of novel ``valleytronic''
applications~\cite{Mak:28,Zeng:29,Xiao:30,Cao:31}. Lastly, carrier confinement
and reduced dielectric screening in these materials leads to large many-body
effects, resulting in bound state complexes of electrons and holes with very
large binding
energies~\cite{Chernikov:32,He:33,Ugeda:34,Berkelbach:9,Mak:17,Shang:35,Ross:36}.
In this work we focus on this latter property, providing a deeper
understanding of the factors that control the binding energies of electron-hole
complexes in two-dimensional TMDCs.

From the computational perspective, the most accurate means of describing
excitonic properties in periodic solids currently available is the GW+BSE
approach~\cite{Hybertsen:37,Rohlfing:38,Chei:39,Rama:40,Qiu:41}. Unfortunately,
analogous fully {\em ab initio} approaches have not been developed for
the treatment of larger electron-hole complexes such as trions and
biexcitons~\cite{Berkelbach:9,Mak:17,Shang:35,Ross:36}.  However, simplified
approaches have proved to be effective, building on well-established
coarse-grained methodologies developed for semiconductor quantum wells and
other nanostructures.  An effective real-space electron-hole potential is
combined with an approximate treatment of the band structure, such as an
effective mass model or a few-band tight-binding model, to build the model
Hamiltonian. This also has roots in the early discussion of the Bethe-Salpeter
approach by Hanke and Sham, demonstrating the relationship to the
phenomenological approach of Wannier~\cite{Hanke:42}.

The pioneering work of Keldysh highlighted the fact that screening effects in
quasi two-dimensional systems are intrinsically non-local~\cite{Keldysh:3}.
Using a generalized Keldysh approach, Cudazzo {\em et al.} formulated a simple and
successful theory for excitons in graphane~\cite{Cudazzo:10,Cudazzo:43}. This
approach has since been applied by various authors to study optical spectra as well
as the properties of excitons, trions, and biexcitons in
TMDCs~\cite{Berkelbach:8,Berghauser:44,Zhang:45,Fogler:46,Wang:47,Wu:48}. These
studies have produced exciton binding energies and real-space structures that
are in reasonable quantitative agreement with first principles GW+BSE
calculations and experiments in a variety of two-dimensional TMDC
systems~\cite{Huser:49}. This fact is not entirely surprising for three
reasons.  First, recent {\em ab initio} calculations show that the effective
quasiparticle interactions that emerge at the RPA level nearly perfectly match
those used to describe the effective electron-hole interaction in the models
mentioned above~\cite{Steinhoff:50}. Second, the {\em ab initio} band structure near the $K$-point
is well described by elementary two- and three-band
models~\cite{Xiao:30,Berkelbach:51}. Third, the spatial extent of the
exciton that emerges from fully {\em ab initio} calculations is sufficiently
large relative to the atomic scale to suggest that a coarse-grained Hamiltonian
is justified~\cite{Ugeda:34}.

Even within the simplified framework of an effective Hamiltonian, the exact
solution of the multi-body Schr\"odinger equation for larger electron-hole
complexes is challenging.  Initial work on exciton and trion binding energies
in TMDCs employed variational wave functions~\cite{Berkelbach:8}. This approach
has been used more recently and with more intricate trial wave functions to
study biexcitons~\cite{Berkelbach:9}.  In both cases the results found from
variational solutions of the effective few-body Schr\"odinger equation are in reasonable
agreement with experimental results.  However, since binding energies for
trions and biexcitons are extracted with reference to the exciton binding
energy, the use of variational wave functions for all excitonic complexes leads
to binding energies that need not provide a lower bound to the ``exact'' value,
and it is unclear how much error cancellation occurs as a result.  For the
trion binding energy, Ganchev {\em et al.} have discovered a remarkable {\em
exact} solution, but only for the case where the full Keldysh effective
potential is replaced with a completely logarithmic form that is accurate only
at short range~\cite{Aleiner:4}. It is the purpose of this work to investigate
the nature and accuracy of these approximate solutions by comparing
with numerically exact results, and thereby to provide insights into the
properties of higher-order excitonic complexes in two-dimensional TMDCs.

\begin{table*}[t!]
\begin{tabular*}{\textwidth}{@{\extracolsep{\fill}} l  c  c  c  c  c  c }
  \hline  
  \hline  
          & DMC X (eV) & variational X & DMC X$^{-}$ (meV) & experimental X$^-$ & DMC XX (meV) & experimental XX\\
  \hline   
  MoS$_2$  & 0.5514 & 0.54 & 33.8 & 43~\cite{Zhang:18}, 18~\cite{Mak:17} & 22.7 & 70~\cite{Mai:15} \\
  MoSe$_2$ & 0.4778 & 0.47 & 28.4 & 30~\cite{Ross:13} & 17.7 & \\
  WS$_2$   & 0.5191 & 0.50 & 34.0 & 30~\cite{Plechinger:16}, 45~\cite{Zhu:56} & 23.3 & 65~\cite{Plechinger:16} \\
  WSe$_2$  & 0.4667 & 0.45 & 29.5 & 30~\cite{Jones:14} & 20.0 & 52~\cite{Berkelbach:9} \\
  \hline  
  \hline  
\end{tabular*}
\caption{Estimated exciton (X), trion (X$^-$), and biexciton (XX) binding
energies for different members of the 2D TMDC class of materials. Where two
numbers are reported, the number on the left is the most current estimate. The
statistical uncertainty in the DMC data is on the order of 0.1--0.3 meV.  The
column labeled `variational' refers to results based on the Keldysh form, taken
from Ref.~\cite{Berkelbach:8}.}
\label{tab:dmc}
\end{table*}

Diffusion Monte Carlo (DMC) provides a useful approach for studying the
energetics of excitonic complexes.  Briefly, the DMC algorithm propagates an
initial wavefunction in imaginary time using a Jastrow-based guiding
wavefunction until the exact ground state wavefunction and energy is obtained.
Technical details of our DMC calculations can be found in the Supplemental
Material.  At convergence, DMC yields numerically exact exciton, trion and
biexciton ground-state energies within the confines of an effective few-body
Schr\"odinger equation.  Specifically, our calculations employ an effective
mass treatment of the band structure and an electron-hole interaction
appropriate for the two-dimensional TMDC family of materials, i.e. 
\begin{equation}
H = -\sum_i \frac{\nabla_i^2}{2m_i} + \sum_{i<j} q_i q_j V(r_{ij})
\end{equation}
where
\begin{equation}
\label{keldysh}
V(r) = \frac{\pi}{(\epsilon_1 + \epsilon_2)r_0}\left[H_0(r/r_0) - Y_0(r/r_0)\right];
\end{equation}
in the potential above, $H_0$ is the Struve function, $Y_0$ is the Bessel
function of the second kind, and $\epsilon_1$ and  $\epsilon_2$ are the
dielectric constants for the material above and below the TMDC layer; in all
results presented, we use $\epsilon_1 = \epsilon_2 = 1$, relevant for `ideal'
or suspended TMDC monolayers.

In addition, DMC allows a full sampling of the square of the wavefunction,
which can be used to extract insight into the structure of small bound carrier
assemblies. Although DMC has previously been used to calculate ground-state
properties for trions interacting with a purely logarithmic
potential~\cite{Aleiner:4}, to the best of our knowledge it has not been used
to calculate trion properties with the more realistic electron-hole
interaction above, nor has it been used to calculate the properties of biexcitons.
It should be noted that while the present work was underway, a numerically
exact finite temperature path integral Monte Carlo (PIMC) study of excitons,
trions, and biexcitons using the full Keldysh effective potential
appeared~\cite{PIMC:11}.  While we believe that DMC is a more direct method
than PIMC for the study of what are essentially ground state properties, we
note that the results presented here are in quantitative agreement with those
presented earlier in Ref.~\cite{PIMC:11}, yielding ground state energies
that lie below those of Ref.~\cite{PIMC:11} by fractions of a percent.
On the other hand, the goals of this work are somewhat distinct from those of
Ref.~\cite{PIMC:11}.  In particular, we focus on the specific physical
factors that influence the delicate balance of relative trion and biexciton
binding energies, as well as the accuracy of variational approaches in light of
the ``exact'' DMC results.

\begin{figure}[b!]
\centering
\includegraphics[scale=1.0]{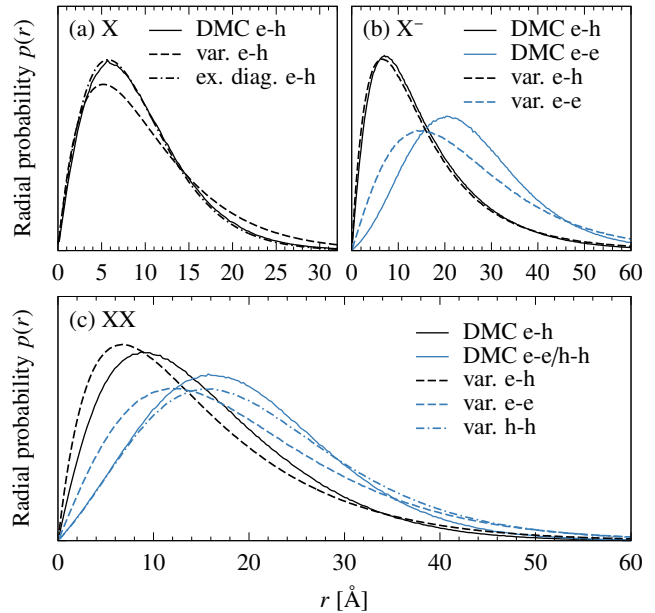}
\caption{(a) Radial probability distributions for the distance $r_{\t{eh}}$ of
an exciton in MoS$_2$; `ex.~diag.' refers to the result of a grid-based exact
diagonalization. (b) Radial probability distributions for the distances
$r_{\t{eh}}$ and $r_{\t{ee}}$ of a negative trion in MoS$_2$. (c) Radial
probability distributions for the distances $r_{\t{eh}}$, $r_{\t{ee}}$, and
$r_{\t{hh}}$ of a biexciton in WSe$_2$. For the
DMC calculation, the curves for $r_{\t{ee}}$ and $r_{\t{hh}}$ coincide because
the electron and hole effective masses are taken to be equal. 
}
\label{fig:keldysh}
\end{figure}

\begin{table*}[t!]
\begin{tabular*}{\textwidth}{@{\extracolsep{\fill}} lccccc ccccc }
  \hline  
  \hline  
          & \multicolumn{5}{c}{Trion} & \multicolumn{5}{c}{Biexciton} \\
  \cline{2-6}\cline{7-11}
          & Keldysh & pure ln & pure $1/r$ & variational 1 & variational 2    &  Keldysh & pure ln & pure $1/r$ & variational 1 & variational 2\\
  \hline   
  MoS$_2$  & 33.8 & 48.9 & 1630 & 26 & 14   & 22.7 & 26.2 & 2610 &         \\
  MoSe$_2$ & 28.4 & 39.3 & 1780 & 21 & 12   & 17.7 & 21.2 & 2820 &    &    \\
  WS$_2$   & 34.0 & 53.6 & 1050 & 26 & 14   & 23.3 & 28.7 & 1680 &    &    \\
  WSe$_2$  & 29.5 & 44.9 & 1120 & 22 & 12   & 20.0 & 23.9 & 1780 & 37 & 16 \\
  \hline
  \hline
\end{tabular*}
\caption{Comparison of trion and biexciton binding energies for several
potential forms in units of meV obtained from DMC, except for the column
labeled `variational'; the latter results are based on the
Keldysh form and taken from Ref.~\cite{Berkelbach:8} for trions and
Ref.~\cite{Berkelbach:9} for the WSe$_2$ biexciton. Binding energies in the
`variational 1' column are with respect to the variational exciton binding
energies, whereas those in the `variational 2' column are with respect to the
exact DMC exciton binding energies.  For the DMC Keldysh and pure logarithmic
potentials, the uncertainty is of the order of 0.1--0.3 meV.  For the pure
Coulombic potential, the uncertainty is of the order of 10 meV.}
\label{tab:pots}
\end{table*}

In Tab.~\ref{tab:dmc}, we report exciton, trion, and biexciton binding energies
calculated via DMC and compare to those extracted in recent experiments.  The
DMC exciton binding energies, defined as $E_{\t{b}}^{\t{X}} = -E^{\t{X}}$, are
only 2--4\% larger than those obtained in previous variational calculations
that employed a trial wavefunction of the form $\Psi_{\t{T,var}}(r_{\t{eh}})
\sim \exp\left(-r_{\t{eh}}/a\right)$. The radial probability distribution for the
distance $r_{\t{eh}}$, which completely determines the exciton wavefunction, is
plotted in Fig.~\ref{fig:keldysh}(a); we compare the variational wavefunction to
results obtained via DMC as well as a grid-based exact diagonalization of the
one-dimensional Sch\"odinger equation.  The simple $1s$-like variational
wavefunction matches the true ground-state wavefunction well, but does not
decay rapidly enough for large $r$. 
As we will show, achieving a similar level of agreement between exact DMC and variational
estimates for the binding energies of larger excitonic complexes is, in
principle, a much more difficult task because trion and biexciton wave
functions are more elaborate and their approximation may in principle require
many variational parameters to achieve a high level of accuracy.  

The DMC trion binding energies given in Tab.~\ref{tab:dmc} are all in the range
of 28--34 meV, which is in excellent agreement with current experimental
estimates; however it should be noted that realistic substrate effects have
been ignored in the present calculations.
In Tab.~\ref{tab:pots}, we compare trion binding energies for two additional potentials: a
purely logarithmic form and an unscreened $1/r$ Coulombic form. These two potentials
represent the asymptotic small and large $r$ behavior, respectively,
of~(\ref{keldysh}). The purely logarithmic potential
approximation has been employed by Ganchev {\em et al.} in their analytical
treatment of trions in TMDCs~\cite{Aleiner:4} while the Coulomb potential is the standard form
for three-dimensional semiconductors. 
We find that a purely logarithmic
potential overbinds the trion and results in binding energies about 50\%
larger than those reported by experiments. Unsurprisingly, the pure Coulombic
potential vastly overbinds the complex, resulting in binding energies that are
30--50 times too large and with a different ordering than is the case for
the full potential~(\ref{keldysh}), which is material dependent.  
Coulombic binding energies would of course be reduced with the inclusion of a
static dielectric constant significantly larger than unity.

Our trion binding energies are about 30\% larger
than those calculated variationally. 
The two-parameter variational trial wavefunction used in the trion calculations
was~\cite{Berkelbach:8}
\begin{equation}
\Psi_{\t{T,var}}(r_{\t{e}_1\t{h}}, r_{\t{e}_2\t{h}}) \sim \exp\left(-r_{\t{e}_1\t{h}}/a-r_{\t{e}_2\t{h}}/b\right) + \{a\leftrightarrow b\},
\end{equation}
\noindent a form inspired by Chandrasekhar's treatment of the hydrogen
anion~\cite{Chandrasekhar:1944}.  Although Fig.~\ref{fig:keldysh}(b) shows that
this optimized variational wavefunction reproduces $p(r_{\t{eh}})$ almost
exactly, the variational form does not capture the electron-electron repulsion
properly because it lacks any explicit $r_{\t{e}_1}-r_{\t{e}_2}$ correlation
terms.  The peak of the electron-electron distribution is at too small a
radius, which results in over-estimating the electron-electron repulsion and
underestimating the trion binding energy, as seen in Tab.~\ref{tab:pots}.
Nonetheless, the level of agreement is surprisingly good given the simplicity
of the variational wave function employed in Ref.~\cite{Berkelbach:8}.
Furthermore, by treating the exciton and trion on equal footing with physically
similar variational wavefunctions, a fortuitous cancellation of total energy
errors leads to binding energies which are quite close to the exact results
(`variational 1' column in Tab.~\ref{tab:pots}); referencing the variational
trion energy to the exact DMC exciton energy (`variational 2' column in
Tab.~\ref{tab:pots}) leads to a significant underestimation of the binding
energy, albeit one that is a genuine lower bound.

\begin{figure}[b!]
\centering
\includegraphics[scale=1.0]{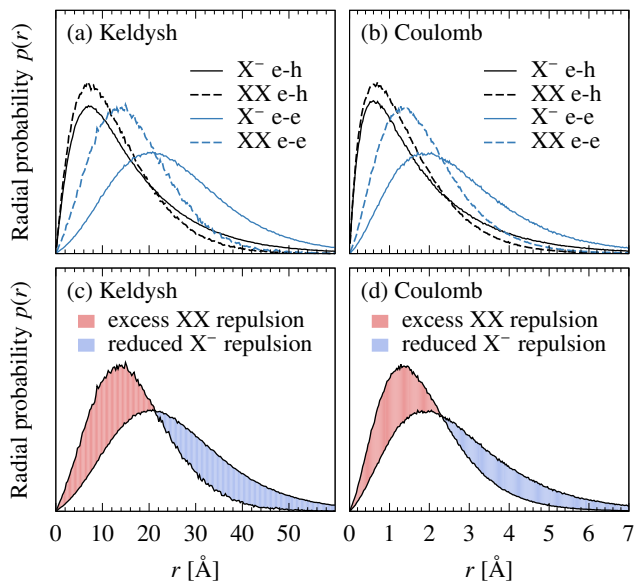}
\caption{(a) Radial probability distributions for the distances $r_{\t{eh}}$ and $r_{\t{ee}}$
in the trion (X$^-$) and biexciton (XX) using a Keldysh form for the inter-carrier
potential. 
(c) The same $r_{\t{ee}}$ distributions as in panel (a), but the relocated
repulsive weight is shaded. 
(b),(d) The same as in panels (a),(c) but using a Coulombic form for the
inter-carrier potential. 
}
\label{fig:pots}
\end{figure}

Finally, in Tabs.~\ref{tab:dmc} and \ref{tab:pots} we report biexciton binding
energies $E_{\t{b}}^{\t{XX}} = 2E^{\t{X}} - E^{\t{XX}}$ , and in
Fig.~\ref{fig:keldysh}(c) we compare carrier probability distributions obtained
from DMC and from a recent six-parameter variational
calculation~\cite{Berkelbach:9}. Other than the $r_{\t{hh}}$ distance, which is
accurately predicted, the variational wavefunction is a bit too compact, leading to a
total variational energy which is slightly too high.  When referenced to the exact
DMC exciton energy, we note that the variational biexciton binding energy is actually
quite accurate -- only 3.5 meV (about 15\%) too small.  When referenced to the
\textit{less accurate} variational exciton energy, the mis-matched cancellation
of errors is such that the biexciton binding energy is slightly
\textit{overestimated}.

In our DMC calculations, we find that the full potential
with dielectric screening~(\ref{keldysh}) yields binding energies that are
smaller than either of the other potentials considered, and that are smaller
than experimental estimates by 60--70\%.  Most significantly, exact DMC calculations show
that the binding energies for biexcitons are significantly \textit{smaller}
than that of trions.  This fact, which has been noted in recent PIMC
calculations as well~\cite{PIMC:11}, disagrees with recent experimental
estimates and is at odds with expectations that emerge from the standard case
of pure Coulombic interactions.	Whereas in the latter $1/r$ case biexcitons are
more strongly bound than trions by a factor of about 1.6, in the purely
logarithmic case the situation is reversed and trions are more strongly bound
than biexcitons.  Interestingly, we find that for realistically parameterized
Keldysh potentials, the biexciton binding energies are slightly smaller than
those found with the purely logarithmic potential, despite the latter being a
presumably ``weaker'' potential; this highlights the subtle balance of energies
involved in the formation of the biexciton.  We note in passing that the
binding energies for biexcitons obtained with the logarithmic potential are
significantly closer to the full Keldysh results than they are for
trions, suggesting that the short-range approximation of Ganchev \textit{et
al.}~may be even better for biexcitons. This result is consistent with the
smaller real-space structure of the biexciton seen by comparing 
Figs.~\ref{fig:keldysh}(b) and (c).

To gain deeper insight into this balance of energies, we consider the electron-hole
and electron-electron distributions for trions and biexcitons obtained with the Keldysh
and Coulombic potentials, plotted in Fig.~\ref{fig:pots}(a),(b).
Suppose that for a given potential, the attractive electron-hole probability
profiles were identical for
the trion and the biexciton, and the repulsive electron-electron (hole-hole)
profiles were also identical for the trion and the biexciton. Then elementary
arguments using the definition of the trion and biexciton binding energies,
along with the pairwise additive potential, show that the biexciton binding
energy would be exactly twice the trion binding energy,
$E_{\t{b}}^{\t{XX}}/E_{\t{b}}^{\t{X}^-} = 2$. Any deviations from this ideal 
ratio are due to \textit{relative} differences in the attractive and repulsive probability
distributions as the second hole is added to the negative trion. 

Instead, biexciton-to-trion binding energy ratios of less than 2 are observed
for both the screened interaction~(\ref{keldysh}) and the Coulomb interaction.
In both cases, about 9\% of the total weight in the $p(r_{\t{eh}})$
(attractive) profile is relocated from long $r$ to short $r$. More
significantly, a much larger fraction of the total weight in the
$p(r_{\t{ee}})$ (repulsive) is relocated to short $r$, leading to a reduced
biexciton-to-trion ratio.  Specifically, for the screened
potential~(\ref{keldysh}), about 31\% of the weight is relocated, which leads
to a biexciton binding energy that is smaller than the trion binding energy;
for the Coulomb potential, only 26\% of the weight is relocated, and the
biexciton binding energy remains larger than the trion binding energy.  The
relocated repulsive area is shaded for both potentials in
Fig.~\ref{fig:pots}(c),(d).  From an energetic standpoint, this more notable
change in the repulsive profile occurs because in the trion, there is no reason
for the like charges to be physically close in space. However, in the
biexciton, the complex can achieve stabilizing electron-hole interactions by
having the like charges closer together in space.

A final question that may be raised concerns the qualitative difference between
biexcitonic stability as found by DMC calculations and that extracted from
experiments. As mentioned above, experimentally reported biexciton binding
energies significantly exceed experimentally determined trion binding energies,
and are about a factor of two or more larger than calculated DMC values.  
Since
our DMC values are exact within the confines of the effective mass and
effective potential models, one possibility is that these model ingredients are
oversimplified, and need to be amended.
Although we have neglected screening from the substrate and surrounding environment,
the results of Ref.~\cite{PIMC:11} suggest that the biexciton binding energy will
remain less than the trion binding energy, in so far as these latter screening
effects can be captured in the effective Keldysh potential.
In particular, it is unclear if the
assumption of effective pair-wise additive interactions is a good one for
larger excitonic complexes; perhaps three-body or higher-order interactions are
needed.  On the other hand, experimental determination of
biexciton binding energies in the TMDCs is quite difficult and involves both
assumptions of the nature of spectral signals as well as extrapolations.
Clearly future work should be devoted to addressing this interesting
discrepancy between theory and experiment.

\vspace{1em}

\textit{Note added--} Since this work was completed, a preprint has appeared
which uses high-accuracy stochastic variational Monte Carlo to calculate
the properties of excitons, trions, and biexcitons in monolayer 
TMDCs~\cite{Zhang:2015}. The results are in agreement with the present manuscript
and the authors further speculate as to the origin of the biexciton discrepancy
noted above.

\vspace{1em}

The authors would like to thank Andrey Chaves and Tony F. Heinz for useful discussions.
MZM is supported by a fellowship from the National Science Foundation
under grant number DGE-11-44155. TCB is supported by the Princeton 
Center for Theoretical Science. Part of this work was done with facilities at the Center 
for Functional Nanomaterials, which is a U.S. DOE Office of Science User Facility, 
at Brookhaven National Laboratory under Contract No. de-sc0012704 (MSH).

\clearpage
\onecolumngrid
\begin{center}
\textbf{\large Supplemental Material:\\ 
Binding energies and spatial structures of small carrier complexes\\
in monolayer transition metal dichalcogenides via diffusion Monte Carlo
}
\end{center}

\setcounter{equation}{0}
\setcounter{figure}{0}
\setcounter{table}{0}
\setcounter{page}{1}
\makeatletter
\renewcommand{\theequation}{S\arabic{equation}}
\renewcommand{\thefigure}{S\arabic{figure}}

\onecolumngrid

\section*{Computational details}
In this section we outline the computational approach and model used to
investigate excitonic complexes. Our technique of choice is diffusion Monte
Carlo (DMC). Because DMC is such a widely used approach, we do not give a
detailed account of the method, and refer the reader to more complete technical
discussions~\cite{supp-Needs:1,supp-Ceperley:52,supp-Foulkes:53}.

DMC calculations use the imaginary time Schr\"odinger equation along with a
guiding wavefunction $\Psi_{\t{T}}$ to project out excited states from an initial
wavefunction $\Phi(t=0)$, propagating it in imaginary time until the true
ground state wavefunction $\psi_0$ is found. If we define the mixed probability
$f(\b{R},t) = \Psi_{\t{T}}(\b{R})\Phi(\b{R},t)$~\cite{supp-Grimm:54,supp-Kalos:55}, where $\b{R}$
contains the spatial coordinates of all quantum particles, then the equation of motion
for $f(\b{R},t)$ derived from the imaginary time Schr\"odinger equation
is
\begin{equation}
-\frac{\partial f(\b{R},t)}{\partial t} = -\frac{1}{2}\nabla_{\b{R}}^2 f(\b{R},t) 
    + \nabla_{\b{R}}\cdot[\b{v}(\b{R})f(\b{R},t)]
    + (E_{\t{L}}(\b{R},t) - E_{\t{ref}})f(\b{R},t),
\end{equation}
where 
\begin{equation}
\b{v}(\b{R},t) = \Psi_{\t{T}}^{-1}(\b{R})\nabla_{\b{R}}\Psi_{\t{T}}(\b{R}),
\end{equation}
\begin{equation}
E_{\t{L}}(\b{R}) = \Psi_{\t{T}}(\b{R})^{-1}\hat{H}\Psi_{\t{T}}(\b{R}),
\end{equation}
and $E_{\t{ref}}$ is an arbitrary energy offset. 
A solution to the importance-sampled imaginary-time Schr\"odinger equation is then
sampled using approximate Greens functions that result in the drift-diffusion motion and the 
branching action~\cite{supp-Needs:1}. For the systems we consider, the exact ground state wave function is nodeless,
so DMC yields {\em exact} ground state energies and a sampling of the {\em
exact} ground state wavefunctions. 

The guiding wavefunction used in this work is of the form $\Psi_{\t{T}}(\b{R})
= e^{J(\b{R})}$, which contains the Jastrow factor introduced in
Ref.~\cite{supp-Aleiner:4} adapted specifically to the
potential~(2). The Jastrow term contains two-body electron-hole and
electron-electron (hole-hole) terms
\begin{equation}
\label{ueh}
u_{\t{eh}}(r) = c_1r^2\ln(r)e^{-c_2r^2} - c_3 r \left(1 - e^{-c_2r^2}\right),
\end{equation}
\begin{equation}
\label{uee}
u_{\t{ee}}(r) = c_4r^2\ln(r)e^{-c_5r^2}.
\end{equation}
The constants $c_1 = m_\t{e}m_\t{h}/2(m_\t{e} + m_\t{h})$ and $c_4 = -m_\t{e}/4$ 
($m_{\t{e,h}}$ are the effective masses of the carriers) were chosen to satisfy
the logarithmic analogue of the Kato cusp conditions; the other constants were optimized via unreweighted
variance minimization to improve the efficiency of the Monte Carlo
sampling~\cite{supp-Umrigar:5,supp-Drummond:6}. A blocking method is used to gauge the
correlation timescales for the energy estimates, and yields accurate standard
deviations for the final average~\cite{supp-Flyvbjerg:12}.
Energy estimates were obtained for calculations with $\Delta t \in
\{0.01,0.03,0.1\}$, and then extrapolated to zero timestep. 
All reported DMC probability distributions
were sampled from forward-walking DMC calculations~\cite{supp-Barnett:7} with the
optimal guiding wave function described by Eqs.~(\ref{ueh})--(\ref{uee}), and
$\Delta t = 0.01$. A forward walking time of 300 a.u.~was used for calculations employing the
Keldysh potential; that time was 30 a.u.~for calculations using the 
Coulomb potential.

Screening lengths and effective masses for all materials studied were
determined in Ref.~\cite{supp-Berkelbach:8}.  Potentials of the type (2) were
first discussed by Keldysh~\cite{supp-Keldysh:3} and have been used to treat
excitonic properties in quasi two-dimensional materials in several recent
studies~\cite{supp-Cudazzo:10,supp-Cudazzo:43,supp-Berkelbach:8,supp-Berghauser:44,supp-Zhang:45,supp-Fogler:46,supp-Wang:47,supp-Wu:48}.
The potential~(2) behaves as $1/r$ at large $r/r_0$, but diverges more weakly
as $\ln ({r/r_0})$ near $r = 0$. The crossover point is related to the
screening distance $r_0 = 2\pi\chi_{\t{2D}}$, where $\chi_{\t{2D}}$ is the
two-dimensional polarizability of the TMDC layer. 
For computational efficiency and
consistency with past variational calculations, we use a modified form of
the effective potential~(2), given by
\begin{equation}
\label{keldysh2}
V^\prime(r) = -\frac{1}{r_0}\left[\ln\left(\frac{r}{r + r_0}\right) + (\gamma - \ln 2)e^{-r/r_0}\right],
\end{equation}
where $\gamma$ is Euler's constant~\cite{supp-Cudazzo:10}. Calculations
with the unaltered potential~(2) typically result in exciton
ground-state energies that are merely 2--3 meV lower than those produced
with~(2).

\end{document}